\documentclass[useAMS]{emulateapj} 
\usepackage{color}
\usepackage{natbib}
\usepackage{amsmath}
\usepackage{epsfig} 
\usepackage{times}

\def\be{\begin{eqnarray}}   \def\ee{\end{eqnarray}}
\def\ben{\begin{eqnarray*}} \def\een{\end{eqnarray*}} 

\def\sec#1{Section~\ref{sec:#1}}
\def\fig#1{Figure~\ref{fig:#1}} 
\def\tab#1{Table~\ref{tab:#1}}
\def\equ#1{Equation~(\ref{equ:#1})}

\definecolor{grey}{rgb}{0.35,0.35,0.35}
\begin{document}
\title {An empirical formula for the distribution function of a thin 
exponential disc}
\author{Sanjib Sharma \& Joss Bland-Hawthorn}
\affil{Sydney Institute for Astronomy, School of Physics, University of Sydney, NSW 2006, Australia}
\begin{abstract}
An empirical formula for a Shu distribution function 
that reproduces a thin disc with exponential surface density 
to good accuracy is presented. 
The formula has two free parameters that specify 
the functional form of the velocity dispersion.
Conventionally, this 
requires the use of an iterative algorithm to produce the correct
solution, which is computationally taxing for applications 
like Markov Chain Monte Carlo (MCMC) model fitting. 
The formula has been shown to work for flat, rising and falling 
rotation curves. Application of this methodology to one of the 
Dehnen distribution functions is also shown. Finally,  an 
extension of this formula  to reproduce velocity dispersion 
profiles that are an exponential function of radius is also presented.
Our empirical formula should greatly aid the efficient comparison of
disc models with large stellar surveys or N-body simulations.
\end{abstract}

\section{Introduction} 
With the advent of large photometric, spectroscopic and proper motion surveys
of stars, it has now become possible to study in detail the properties of the  
Milky Way disc. Substantial improvements in the richness and quality offered
by these new surveys
demands increasing sophistication in the methods used to analyze them
\citep{2012MNRAS.419.2251M,2011MNRAS.413.1889B}. Simultaneously,
the vast increase in the {\it size} of the new data sets requires that these
same methods have greatly enhanced computational efficiency compared
to existing algorithms.
There is a pressing need for wholesale improvements to 
theoretical modelling in order to cope with the data deluge. 
In the context of modelling the kinematics of the disc,
for simplicity, 3D Gaussian functions have been used. This
form is barely adequate for radial and vertical motions and entirely
inappropriate for azimuthal motions. The azimuthal distribution of stellar motions,
the subject of this paper, is very skewed because the surface density and 
radial velocity dispersions are declining functions of radius $R$. 
At any given radius, we observe more stars with azimuthal velocity 
$v_{\phi}$ less than the local circular velocity $v_{\rm c}$ than
stars with $v_{\phi}>v_{\rm c}$, a phenomenon known as `asymmetric
drift' \citep[e.g.,][]{1969ApJ...158.1115S}. The inadequacy of the 
Gaussian functions in modelling disc kinematics has been discussed by
 \citet{2010MNRAS.401.2318B} and \citet{2012MNRAS.419.1546S}.

A better way to model disc kinematics is to use a proper distribution 
function $f$. 
For axisymmetric systems with potential $\Phi=\Phi(R)$, the 
distribution function from Jeans' theorem is a function of energy 
$E$ and angular momentum $L$ alone, $f=f(E,L)$. However, there 
is no unique solution for a given $\Sigma(R)$
\citep{1976ApJ...205..751K}. Specifying $\sigma_R(R)$ can restrict 
the solution space but does not remove the general degeneracy,
the reason being that a finite set of functions of 
one variable cannot determine 
a function of two variables \citep{1999AJ....118.1201D}. 
One approach to arrive at a suitable $f(E,L)$ as given by
\citet{1969ApJ...158..505S} is to consider moderately
heated (`warmed up') versions of cold discs, i.e.
\be
f(E,L)=\frac{F(L)}{\sigma_R^2(L)}{\rm exp}\left[-\frac{E-E_c(L)}{\sigma_R^2(L)}\right]
\ee
where $E_c(L)$ is the energy of a circular orbit with angular momentum $L$.
Here $F(L)$ is chosen such that in the
limit $\sigma(L) \rightarrow 0$, the function gives a surface density of 
$\Sigma(R)$. Recently \citet{2012MNRAS.419.1546S} used this
to derive important formulas for disc kinematics.

Forms of warm discs other than the Shu family can be
found in \citet{1999AJ....118.1201D}. Further examples include
the distribution functions for a quasi-isothermal disc described 
in \citet{2010MNRAS.401.2318B,2012MNRAS.426.1328B}.
One problem with these warmed up functions is
that they reproduce the target density only very approximately $-$ the 
larger the value of $\sigma_R$, the larger the discrepancy. One way to
improve this is to iteratively solve for $F(L)$
from the integral equation connecting $\Sigma(R)$ and $f(E,L)$ 
as suggested by \citet{1999AJ....118.1201D}. 
For applications like MCMC model fitting, where in each
iteration the model parameters are changed and one has to 
recompute the solution (Sharma et al 2013, in preparation), existing
algorithms are simply too inefficient.
Hence, in this paper we attempt to find an empirical formula for 
$F(L)$ that reproduces a disc with exponential surface density. 
The form of $F(L)$ in general depends on the choice of $\sigma_R(L)$, 
so to proceed we assume $\sigma_R(L)$ to be an exponential 
function with two free parameters $a_0$ and $q$ 
and then find an empirical formula for $F(L|a_0,q)$.
The formula is shown to work for rotation
curves whose dependence with radius is flat, rising and falling.
Additionally we also 
show that a similar formula can also be derived for one of the 
distribution function proposed by \citet{1999AJ....118.1201D}.
Finally, we present an extension of the formula for the case of 
velocity dispersion profiles that are exponential in radius.

The paper is organized as follows. In Section 2, we review the basic 
theory and the method to numerically compute the distribution
function. Using this in Section 3, we determine an empirical formula 
for the distribution function. 
In Section 4, we analyze the accuracy of our results and compare 
it with alternate solutions. Finally, in Section 5, we conclude and 
summarize our findings.

\section{Theory} \label{sec:theory}
In this paper, we consider two types of distribution
functions for warm discs such that
\be
f_{\rm Shu} & =
&\frac{F(L)}{\sigma^2_R(L)}\exp\left(-\frac{E-E_c(L)}{\sigma^2_R(L)}\right)
\ {\rm and} \\
f_{\rm Dehnen,a} & = &\frac{F(E)}{\sigma^2_R(E)}\exp\left(-\frac{E-E_c(L)}{\sigma^2_R(E)}\right).
\ee
Here $E_c(L)$ is the energy of a circular orbit with angular momentum
$L$ and $E-E_c(L)$ is the energy in excess of that required 
for a circular orbit. Let $R_g$ and $R_E$ be the radius of a 
circular orbit with angular 
momentum $L$ and energy $E$ respectively, which in general 
we denote by $R_c$.
Our aim is to compute $F(R_c)$  that produces a disc 
with an exponential surface density
profile, $\Sigma(R)$. For this, either $\sigma_R(R_c)$ 
has to be specified or  $\sigma_R(R)$. If $\sigma_R(R)$ 
is specified then we also need to solve for $\sigma_R(R_c)$ 
in addition to $F(R_c)$. 
We reduce the velocity dispersion to dimensionless form by 
defining $a=\sigma_R/v_{\rm  circ}$, where 
$ v_{\rm  circ}$ is the circular velocity.
For the remainder of the paper, we assume $a$ to be an exponentially decreasing
function of $R_c$(or $R$) and specify it as follows
\begin{eqnarray}
a & = &a_0 e^{-q R_c/R_d}   
\end{eqnarray}
The choice of the functional form is motivated by the desire to 
produce discs in which 
the scale height is independent of radius. 
For example, under the epicyclic approximation, if $\sigma_z/\sigma_R$
is assumed to be constant, then 
the scale height is independent of radius for $q=0.5$ 
\citep{1982A&A...110...61V}.

Instead of solving for  $F(R_c)$ directly, we proceed by 
solving for $\Sigma(R_c)$, which is the surface density in 
$R_c$ space and satisfies  $\int \Sigma(R_c)\: 2 \pi R_c\: dR_c =1$. 
The distribution function is assumed to be normalized 
such that $\int \int \int f(E,L)\: 2\pi R\: dR\: d v_{\phi}\: d v_{R}=1$. 
For a given distribution function let $P(R,R_c)$,
be the probability distribution in $(R,R_c)$ space.  
Using the relation $2 \pi R_c \Sigma(R_c)=\int P(R,R_c) dR$ 
one can then write $F(R_c)$ in terms of $\Sigma(R_c)$. 
Integrating $P(R,R_c)$ over $R_c$ one gets the integral equation 
connecting $\Sigma(R)$ to $\Sigma(R_c)$ and $a(R_c)$.  
This is the main equation that needs to be solved to determine 
$\Sigma(R_c)$. 
The integral equation for $f_{\rm Shu}$ has
already been discussed in
\citet{2009MNRAS.396..203S,2012MNRAS.419.1546S};
below we provide an 
alternate derivation, both for the sake of completeness and 
the need to generalize it for arbitrary rotation curves. 
A detailed derivation for $f_{\rm Dehnen,a}$ is given in the
appendix.


\subsection{The Shu distribution function}
The energy $E$ of an orbit 
in potential $\Phi(R)$ with angular momentum $L$ is given by 
\be
E=\frac{1}{2}v_R^2+\frac{L^2}{2R^2}+\Phi(R)
\ee
If $R_g(L)$ is the guiding radius which is defined as the radius of a 
circular orbit with angular momentum $L$, then the energy of a circular orbit 
can be expressed as 
\be
E_c(L)=\frac{L^2}{2R_g^2}+\Phi(R_g)
\ee
Defining 
\be
\Phi_{\rm eff}(R,L) &= & \frac{L^2}{2R^2}+\Phi(R) 
\ee
we can write
\be
E-E_c(L) &= & \frac{1}{2}v_R^2+\Phi_{\rm eff}(R,L)-\Phi_{\rm eff}(R_g,L) \\
 & = & \frac{1}{2}v_R^2+\Delta \Phi_{\rm eff}(R,L) 
\ee

We now show how to compute $\Sigma(R)$ for a given  $\Sigma(R_g)$.
For $f_{\rm Shu}$, one can write the probability distribution in $(R,R_g,v_R)$
space as
\be
P(R,R_g,v_R)dR\ dv_R\ dR_g &= &2\pi f(E,L)dL\ dR\ dv_R \\
P(R,R_g,v_R) &= &2\pi f(E,L) \frac{dL}{dR_g} dv_R 
\ee
Using $E-E_c(L)= \frac{1}{2}v_R^2+\Delta \Phi_{\rm eff}(R,R_g)$,
$dL/dR_g=2v_{\rm circ}(R_g)/\gamma^2(R_g)$, $\gamma^2(R)=2/(1+d\ln v_{\rm circ}(R)/d\ln dR)$ and integrating over $v_R$ one gets
\be
P(R,R_g) &= &(2\pi)^{3/2} \frac{F(R_g)}{\sigma_R(R_g)}
\frac{2v_{\rm circ}(R_g)}{\gamma^2(R_g)} \times \nonumber \\
& & \exp\left(-\frac{\Delta \Phi_{\rm eff}(R,R_g)}{\sigma^2_R(R_g)}\right)  
\ee
If $\lim_{r \to \infty} \phi(R)=0$ then there is also an additional
factor of ${\rm erf}\left(\sqrt{\frac{-\phi_{\rm
        eff}(R,R_g)}{\sigma^2_R(R_g)}}\right)$. For the cases
considered in this paper $\lim_{r \to \infty} \phi(R)=\infty$ 
and this factor is unity. 
We now define $a=\sigma_R(R_g)/v_{\rm circ}(R_g)$ and 
\be
K(R,R_g)=\exp\left(-\frac{\Delta \Phi_{\rm
    eff}(R,R_g)}{\sigma^2_R(R_g)}\right)
\ee
then we get
\be
P(R,R_g)  &= &(2\pi)^{3/2} \frac{2 F(R_g)}{\gamma^2(R_g) a(R_g)} K(R,R_g) 
\ee
Using $\int P(R,R_g) dR= 2\pi R_g \Sigma(R_g)$ one can write
\be
F(R_g)=\frac{\gamma^2(R_g)a(R_g) \Sigma(R_g)}{\sqrt{2\pi}g_K(a,R_g)}
\ee
and
\be
P(R,R_g) &= & \frac{2 \pi \Sigma(R_g)}{g_K(a,R_g)} K(R,R_g)
\ee
where 
\be
g_K(a,R_g)=\frac{1}{R_g} \int K(R,R_g) dR
\ee
Finally, the integral equation connecting $\Sigma(R)$ and $\Sigma(R_g)$ 
is given by
\be
\Sigma(R)=\frac{1}{R} \int \frac{\Sigma(R_g)}{g_K(a,R_g)} K(R,R_g) dR_g
\ee

For $v_{\rm circ}(R)=v_c (R/R_0)^{\beta}$, $g_K(a,R_g)$ is independent of
$R_g$. For flat rotation curve, the integral has an analytic solution 
and has been given by \citet{2012MNRAS.419.1546S} as
\ben
g_K(a) & = & \frac{e^c\Gamma(c-1/2)}{2c^{c-1/2}} \ {\rm with}\ c=\frac{1}{2a^2}\\
& \approx & \sqrt{\frac{\pi}{2(c-0.913)}}  \ {\rm for\ large\ c}. 
\een
For non-flat rotation curves, this has to be evaluated numerically.


\subsection{The Dehnen distribution function}
Three variants of the Shu distribution function were proposed by 
\citet{1999AJ....118.1201D} out of which we consider the following case
\be
f_{\rm Dehnen,a}=\frac{F(R_E)}{\sigma_R^2(R_E)}\exp\left(\frac{E-E_c(L)}{\sigma_R^2(R_E)}\right) .
\ee
The integral equation connecting $\Sigma(R)$ and $\Sigma(R_E)$ for
the case of flat rotation curve is given by (see Appendix for derivation)
\be
\Sigma(R)=\frac{1}{R} \int \frac{\Sigma(R_E)}{g_K(a,R_E)} K(R,R_E) dR_E
\ee
with $a=\sigma_R(R_E)/v_{\rm circ}(R_E)$ and 
\be
K(R,R_E)=(R/R_E)^{(1+\frac{1}{a^2})}(1-2\ln(R/R_E))^{\frac{1}{2a^2}}
\ee
and
\ben
g_K(a) & = & \frac{1}{R_E} \int K(R,R_E) dR \\
& = & \left(\frac{e}{1+c}\right)^{1+c} \frac{\Gamma(1+c)}{2} \ {\rm with}\ c=\frac{1}{2a^2}\\
& \approx & \sqrt{\frac{\pi}{2(c+1)}} \ {\rm for\ large\ c}. 
\een


\subsection{The iterative method for solving the integral equation of
  surface density}
In general if we have 
\be
P(R,R_c) &= & \frac{2 \pi \Sigma(R_c)}{g_K(a,R_c)} K(R,R_c).
\ee
Then, as shown in previous section, the integral equation 
that connects $\Sigma(R)$ to $\Sigma(R_c)$ is
\be
\Sigma(R)& = & \frac{\int P(R,R_c) d R_c}{2 \pi R} \nonumber \\
& = & \frac{1}{R} \int \frac{\Sigma(R_c)}{g_K(a,R_c)} K(R,R_c) d
R_c.
\label{equ:integralequ1}
\ee
Also, the equation connecting $\sigma_R(R)$ to $\sigma_R(R_c)$ is given by 
\be
\sigma_R(R) & = & \frac{\int P(R,R_c) \sigma_R^2(R_c) d R_c}{\int
  P(R,R_c) d R_c} \nonumber \\
     & = & \frac{1}{R \Sigma(R)} \int \frac{\Sigma(R_c) \sigma^2_R(R_c)}{g_K(a,R_c)} K(R,R_c) d
R_c.
\label{equ:integralequ2}
\ee
In this paper, we are concerned with the following two cases.    
\begin{itemize}
\item For a given $\Sigma(R)$ and $\sigma_R(R_c)$ solve for 
$\Sigma(R_c)$ and $\sigma_R(R)$.
\item For a given $\Sigma(R)$ and $\sigma_R(R)$ solve for 
$\Sigma(R_c)$ and $\sigma_R(R_c)$.
\end{itemize}
An iterative Richardson-Lucy type algorithm to solve for the above mentioned cases was given by
\citet{1999AJ....118.1201D} and is as follows. 
\begin{enumerate}
\item Start with $\Sigma(R_c)$=$\Sigma(R)$.
If the velocity dispersion also needs to be constrained, set
$\sigma_R(R_c)$=$\sigma_{R}(R)$.
\item Solve for the current surface density $\Sigma'(R)$ and
  velocity dispersion  $\sigma'_R(R)$ profiles.
\item Set $\Sigma(R_c)=\Sigma(R_c)\frac{\Sigma(R)}{\Sigma'(R)}$.  If the velocity dispersion also needs to be constrained, set $\sigma_R(R_c)=\sigma_R(R_c)\frac{\sigma_{R}(R)}{\sigma'_R(R)}$.
\item If the answer is within some predefined tolerance then exit  
else go to step 2.
\end{enumerate}

The above algorithm is easy to setup but non trivial to tune 
for efficiency and accuracy. 
We now discuss the technical details of our 
implementation. We represent the profiles by 
linearly spaced arrays with spacing $\Delta R$ and range 
$R_{\rm min}=\Delta R$ to $R_{\rm max}$.
We now determine the optimum step size to compute the integrals.
At a given $R$, the typical width of the kernel $K(R,R_c)$ is given by 
$R a_0 \exp(-q R/R_d)$. Assuming at least $N$ points within 
this width, the criteria for step size becomes 
\be
\Delta R_{\rm int} & < & R_{\rm min} a_0 \exp(-q R_{\rm min}/R_d)/N \\
\Delta R_{\rm int} & < & R_{\rm max} a_0 \exp(-q R_{\rm max}/R_d)/N, 
\ee
The choice of $N$ in general should be greater than 1.
From the above equation, it can be seen that small $a_0$, 
large $q$, large $R_{\rm max}$ and small  $R_{\rm min}$, 
make the step size small and the task of obtaining 
the solution computationally expensive. 
At a given $R$, the contribution to the integral from 
$R_c>R(1+5 a_0 \exp(-q R_{\rm min}/R_d))$ is very small. 
So the upper range of the integral is comfortably known. But the lower range 
is difficult to determine as $\Sigma(R_c)$ increases with
decrease in $R_c$. At large $R$, we find that 
there can be a non negligible contribution from 
$R_c<R(1-5 a_0 \exp(-q R_{\rm min}/R_d))$. To overcome this,  
in general we integrate on a linearly spaced 
array with spacing $\Delta R_{\rm int}$ and range $\Delta R_{\rm int}$
to $2 R_{\rm  max}$. A choice of $N=8$ was found to be satisfactory. 
Additionally, we find that to avoid numerical ringing artifacts, the integral
in step 2 should be done with step size 
$\Delta R_{\rm int}=\Delta R/2^{\alpha}$ , where $\alpha \geq 0$.

\section{Determining the empirical formula}
In the previous section, we have shown how to compute $\Sigma(R_c)$ 
given a $\Sigma(R)$ and $\sigma_R(R_c)$. 
Let us assume that the target surface density is given by 
\be
\Sigma(R)=\frac{1}{2 \pi R_d^2} e^{-R/R_d}  
\ee
and the velocity dispersion profile satisfies
$\sigma_R(R_c)=v_{\rm circ}(R_c)a_0 \exp(-q R_c/R_d)$,  
where $a_0$ and $q$ are two free parameters.
Now starting with a density 
\be
\Sigma(R_c)=\frac{1}{2 \pi R_d^2} e^{-R_c/R_d}  
\ee
and a given $a_0$ and $q$, we can solve the integral equation  
iteratively using the formalism of \citet{1999AJ....118.1201D}, 
and  compute the required $\Sigma(R_c)$.
Let us denote the difference of the obtained solution from the 
initial guess by
\be
\Delta \Sigma(R_c,R_d,a_0,q) & = & \frac{e^{-R_c/R_d}}{2 \pi R_d^2}-\Sigma(R_c,R_d,a_0,q).
\ee
So, the iterative solution gives us the correction factor $\Delta
\Sigma$ but it will be different for different values of $a_0$ and
$q$.  But since we cannot write the correction factor as an analytic function of
$R_c,R_d,a_0$ and $q$, we must try to derive an empirical function. To this end, we first 
compute $\Delta \Sigma$ for a range of values of $a_0$ and $q$ 
and then analyze them. We first study the distribution function 
$f_{\rm Shu}(E,L)$ for the case of the flat rotation curve and then  
generalize our results for non-flat rotation curves. 
Next, we apply our methodology to a distribution 
function other than $f_{\rm Shu}(E,L)$, namely  $f_{\rm
  Dehnen,a}(E,L)$, but study only the case for flat rotation.
Finally, we show how the formula can be extended for the case of 
velocity dispersion profiles that are exponential in radius.

\begin{table}
\caption{\label{tab:tb1} 
Empirically derived parameters for $R^{\rm max}(q)=c_1 R_d/(1+q/c_2)$ and $\Delta\Sigma^{\rm max}(a_0)=c_3 a^{c_4}_{0}/R_d^2$} 
\centering
\begin{tabular}{|l|l|l|l|l|l|l|} \hline
$f(E,L)$ & Potential $\Phi(R)$ & $\beta$ &c1 & c2 & c3 & c4 \\ \hline 
$f_{\rm Shu}$ & $v_c^2 \ln(R/R_d)$ & &3.740 & 0.523& 0.00976& 2.29\\ \hline
$f_{\rm Shu}$ & $v_c^2 \frac{(R/R_d)^{2\beta}}{2\beta}$& 0.2 & 3.822& 0.524&0.00567 & 2.13 \\ \hline
$f_{\rm Shu}$  & $v_c^2\ln(R/R_d)+$& -0.5 & 3.498& 0.454&
  0.01270& 2.12 \\ 
  &  $v_c^2\frac{(R/R_d)^{2\beta}}{2\beta}$&  & & & & \\ \hline
$f_{\rm Dehnen,a}$ & $v_c^2 \ln(R/R_d)$ & &4.876 & 0.661& 0.00062& 1.62\\ \hline
\end{tabular}
\end{table}

\begin{figure}
 \centering \includegraphics[width=0.48\textwidth]{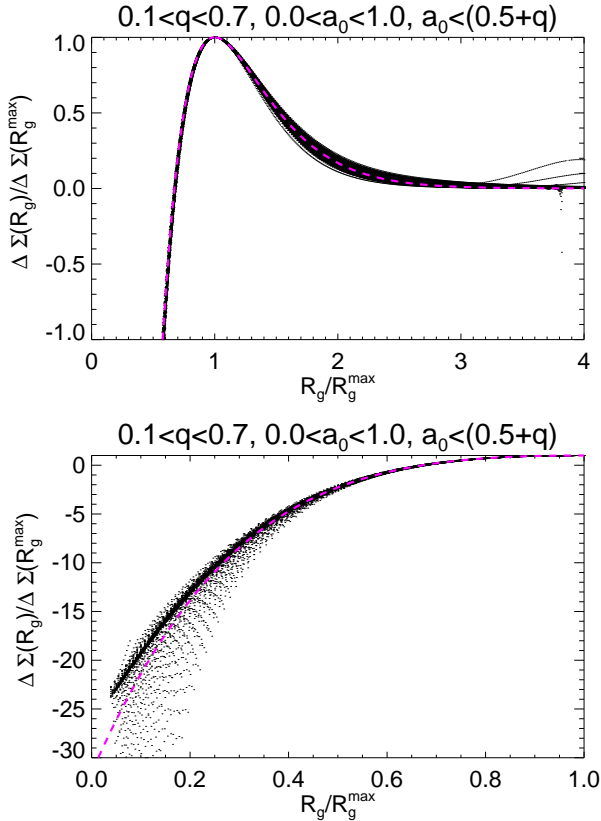}
\caption{ The correction factor $\Delta
  \Sigma$ that is required to reproduce an exponential disc 
  as a function of $R_g$.  $R_g^{\rm max}$ is the radius at which $\Delta
  \Sigma$ is maximum. 
  Different curves in the figure correspond to different values of
  $a_0$ and $q$. The magenta dashed lines correspond to an analytic fit.   
  The bottom panel shows the same set of curves 
  as the top panel but shows the region $0<R_g/R_g^{\rm max}<1$ in 
  greater detail. 
\label{fig:f1}}
\end{figure}

\subsection{Case of $f_{\rm Shu}$ with a flat rotation curve}
The correction factor $\Delta \Sigma$ as a function of $R_g$ for various different 
values of $a_0$ and $q$ is shown in \fig{f1}.
In general, we find that $\Delta \Sigma$ is negative for small $R_g$; 
it then rises to 
a maximum and  then for large $R_g$ it asymptotically approaches
zero (black lines in \fig{f1}). 
If there exists a unique functional 
form, then we must first reduce $\Delta \Sigma$ to a scale invariant
form. To this end, we compute $s(R_g/R_g^{\rm max})=\Delta \Sigma(R_g)/\Delta
\Sigma(R_g^{\rm max})$, $R_g^{\rm max}$ being the value of $R_g$ where
$\Delta \Sigma$ is maximum.  
In \fig{f1}, we plot  $\Delta \Sigma(R_g)/\Delta \Sigma(R_g^{\rm
  max})$ for various different values of $a_0$ and $q$ as derived 
by the iterative solution. It can be seen that they almost follow a 
unique functional form, except for $R_g/R_g^{\rm max}<0.2$, where 
slight differences can be seen (see lower panel \fig{f1}). 
Next we fit this by a function of the following form
\be
s(x) & = &k e^{-x/b}((x/a)^2-1).
\ee
Imposing the condition that $x_{\rm max}=1$, $\int_{-\infty}^{\infty}
x f(x)=0$ and $s(x^{\rm max})=1$, one can solve for $a,b$ and $k$. 
The final function is then given by
\be
s(x) & = &31.53 e^{-x/0.2743}(x/0.6719)^2-1)
\ee 
and this is plotted as magenta dashed lines in \fig{f1}. 
It can be seen that the proposed functional form provides 
a good fit.  For greater accuracy one should replace this 
with a numerical function.

\begin{figure}
 \centering \includegraphics[width=0.48\textwidth]{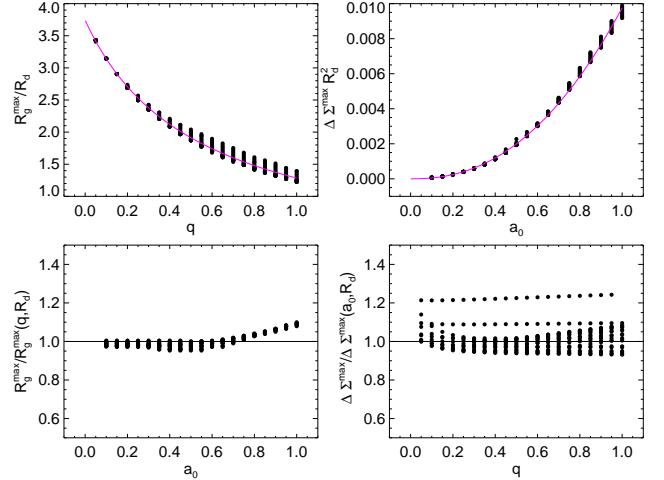}
\caption{ The dependence of $R_{g}^{\rm max}$ and $\Delta
  \Sigma^{\rm max}$ on $q$ and $a_0$. The magenta lines are the 
fitted functional forms  for $R_{g}^{\rm max}$ and $\Delta
  \Sigma^{\rm max}$. The bottom panels demonstrate that 
$R_{g}^{\rm max}$ and $\Delta  \Sigma^{\rm max}$ are nearly
independent of $a_0$ and $q$ respectively. 
\label{fig:f2}}
\end{figure}
We now empirically determine the dependence of $R_{g}^{\rm max}$ and $\Delta
\Sigma(R_g^{\rm max})$ on $R_d,q$ and $a_0$ and this is given below.
\be
\label{equ:rdelta1}
 R_{g}^{\rm max} & = & \frac{c_1}{1+q/c_2}R_d  \\
\Delta \Sigma(R_{g}^{\rm max}) & = & \frac{1}{R_d^2} c_3 a^{c_4}_{0}.
\label{equ:rdelta2}
\ee
with $c_1=3.74$, $c_2=0.523$, $c_3=0.00976$, $c_4=2.29$, 

It was found that the position of maximum $R_g^{\rm max}$ is mainly 
determined by $q$, while the amplitude $\Delta \Sigma(R_g^{\rm max})$
is mainly governed by $a_0$. 
In \fig{f2}, we plot $R_{g}^{\rm max}$ and $\Delta
  \Sigma(R_{g}^{\rm max})$ as functions of $q$ and $a_0$ respectively. 
The fitted functional forms \equ{rdelta1} and \equ{rdelta2} 
are shown as magenta lines.  
The bottom panel shows the dependence of $R_{g}^{\rm max}$ and $\Delta
  \Sigma(R_{g}^{\rm max})$ as a function of $a_0$ and $q$ respectively 
after dividing by the derived functional forms. It is clear 
that there is very little dependence of $R_{g}^{\rm max}$ on 
$a_0$ or of $\Delta  \Sigma(R_{g}^{\rm max})$ on $q$.

The final functional form for $\Sigma(R_g)$ can now be written as
\be
\Sigma_{\rm corr}(R_g) &= & \frac{e^{-R_g/R_d}}{2\pi R_d^2} \nonumber \\
& & -\frac{0.00976 a^{2.29}_{0}}{R_d^2} s\left(\frac{R_g}{3.74R_d(1+q/0.523)}\right)
\label{equ:density_formula}
\ee

\subsection{Case of $f_{\rm Shu}$ with a non-flat rotation curve}
\begin{figure}
 \centering \includegraphics[width=0.48\textwidth]{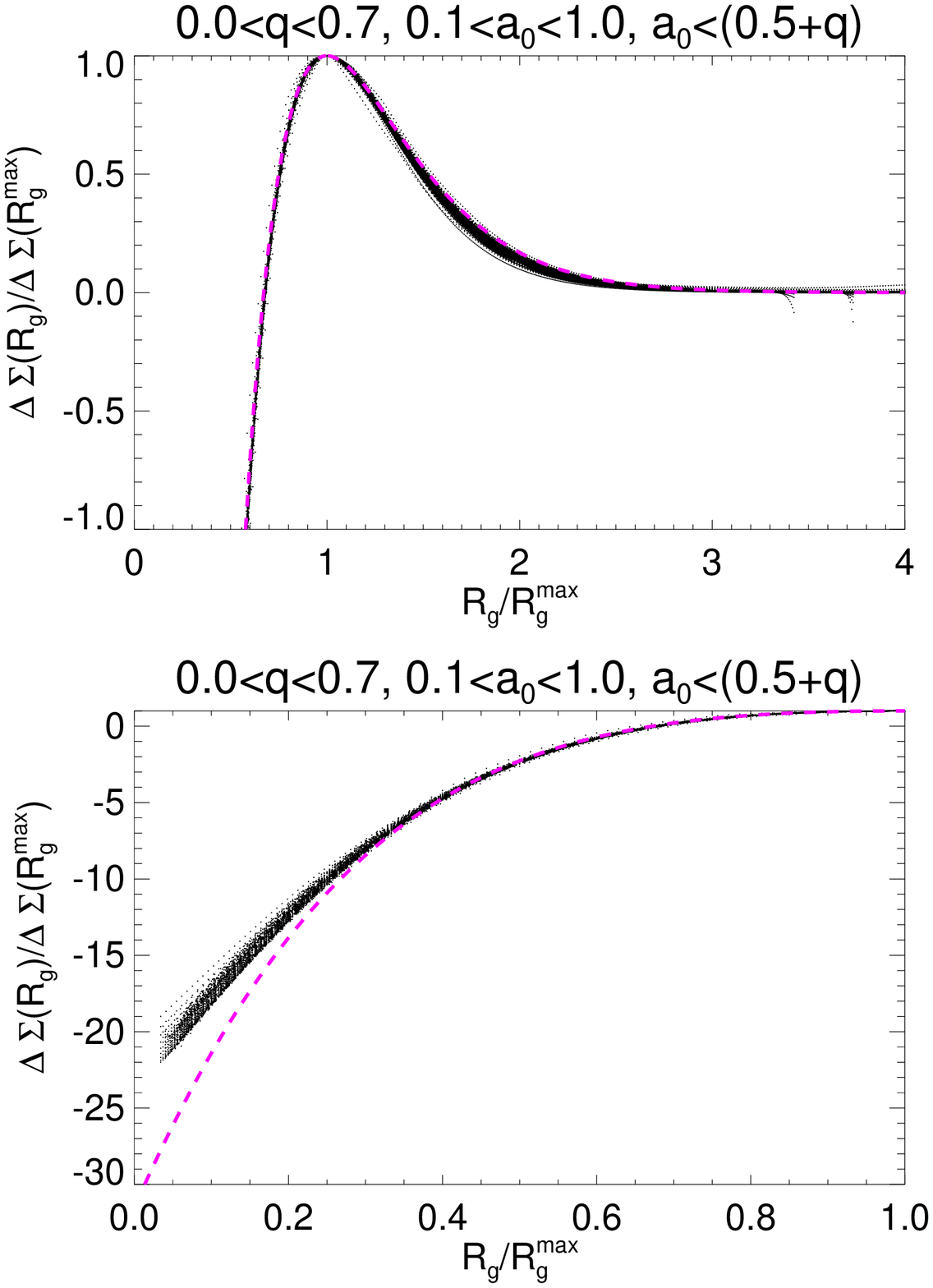}
\caption{ The correction factor $\Delta
  \Sigma$ for distribution function $f_{\rm Shu}(E,L)$ with
  $v_{\rm circ}(R)=v_c (R/R_d)^{0.2}$. For
  description see Figure1.
\label{fig:f3}}
\end{figure}

\begin{figure}
 \centering \includegraphics[width=0.48\textwidth]{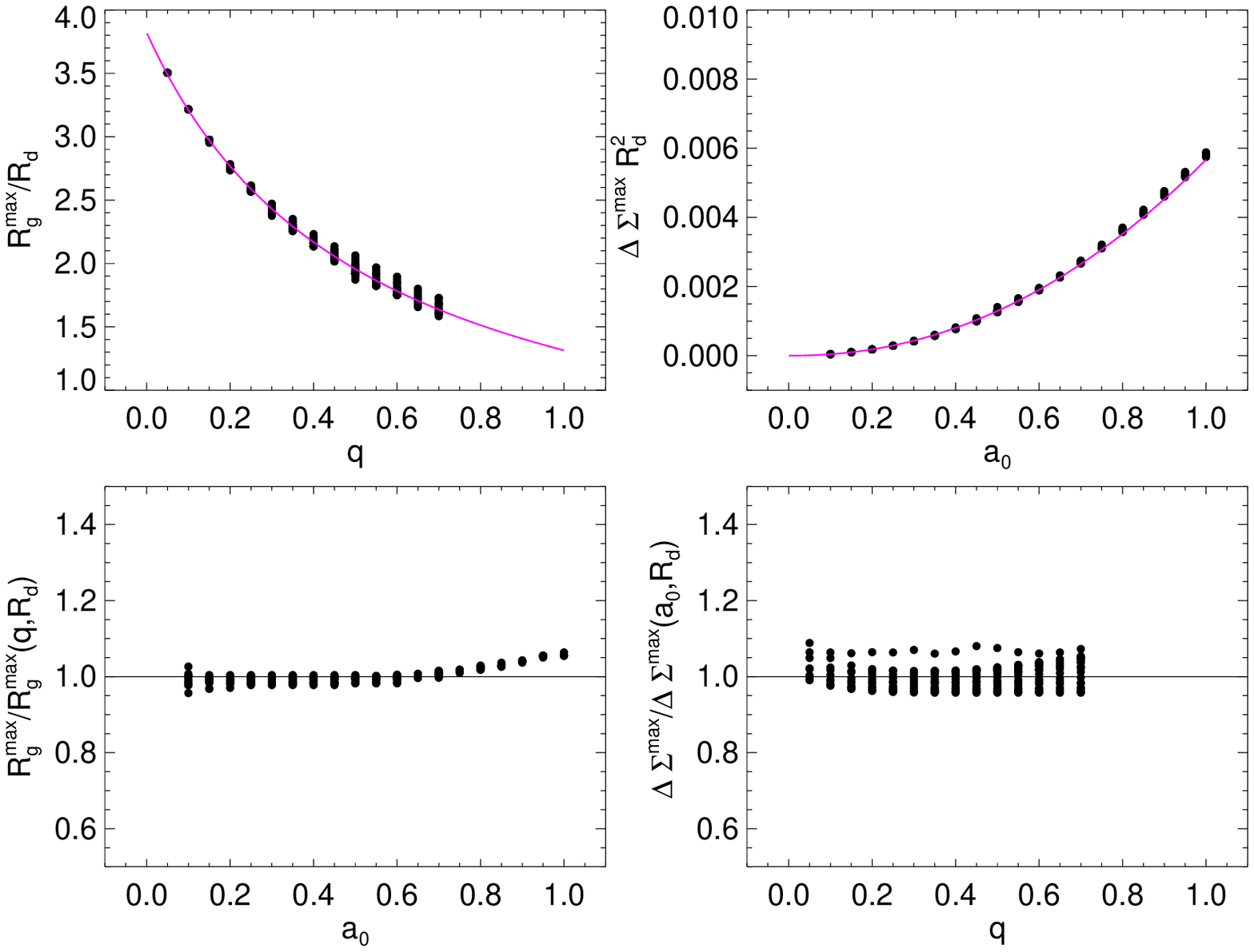}
\caption{ The dependence of $R_{g}^{\rm max}$ and $\Delta
  \Sigma^{\rm max}$ on $q$ and $a_0$ for distribution function $f_{\rm
    Shu}(E,L)$ with  $v_{\rm circ}(R)=v_c (R/R_d)^{0.2}$. For
  description see Figure2.
\label{fig:f4}}
\end{figure}

\begin{figure}
 \centering \includegraphics[width=0.48\textwidth]{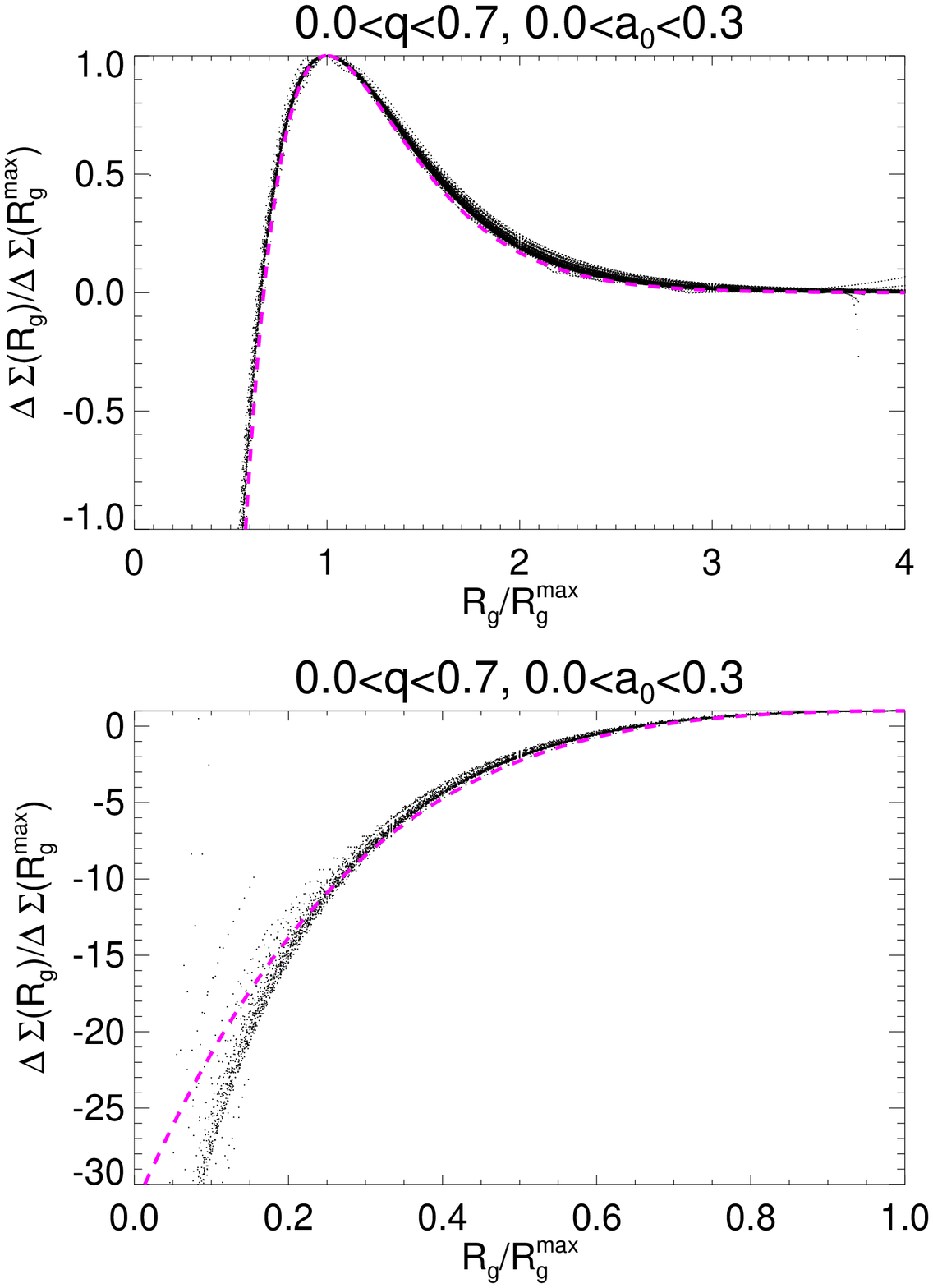}
\caption{ The correction factor $\Delta
  \Sigma$ for distribution function $f_{\rm Shu}(E,L)$ with
  $v_{\rm circ}(R)=v_c\sqrt{1+ R_d/R}$. For
  description see Figure1.
\label{fig:f5}}
\end{figure}

\begin{figure}
 \centering \includegraphics[width=0.48\textwidth]{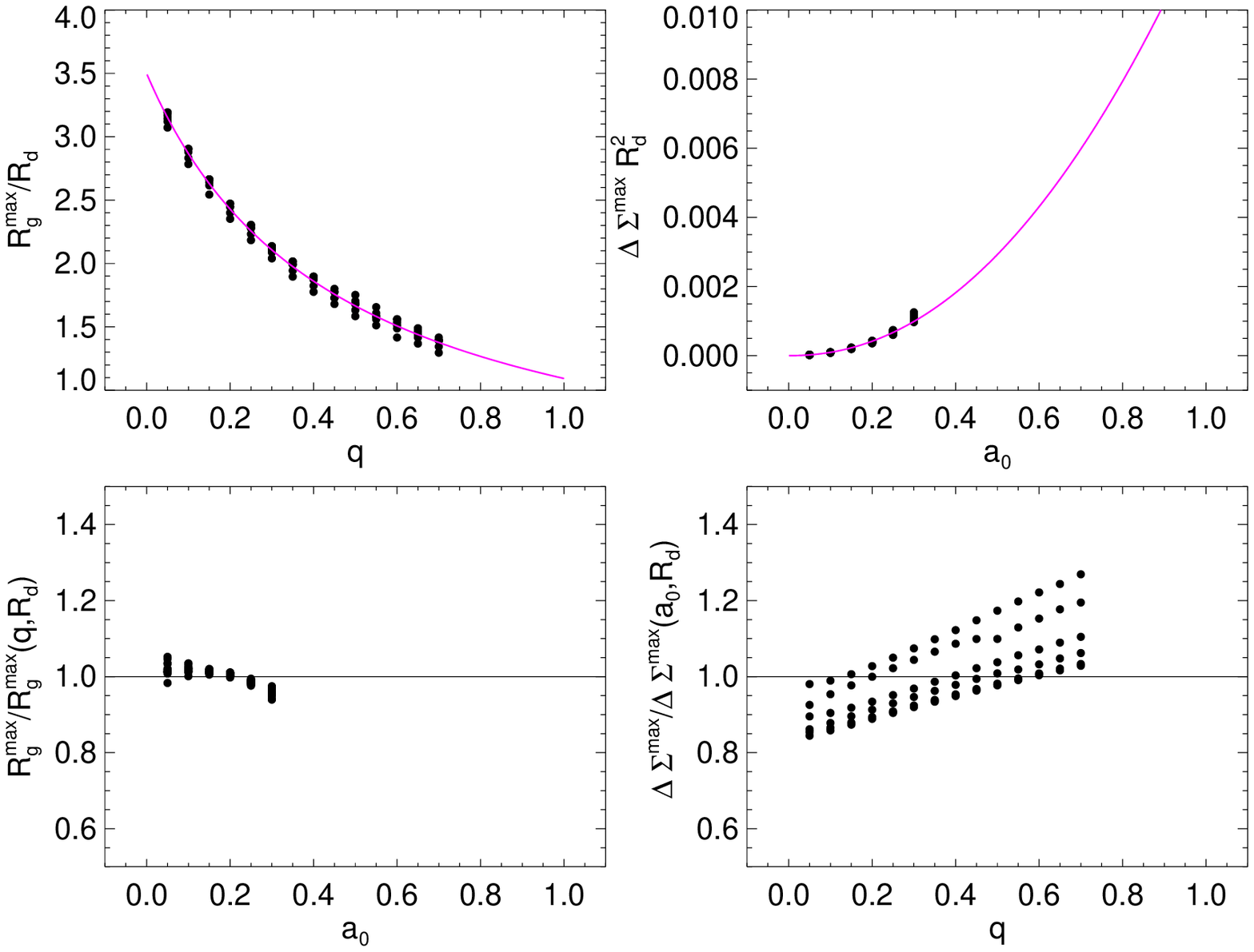}
\caption{ The dependence of $R_{g}^{\rm max}$ and $\Delta
  \Sigma^{\rm max}$ on $q$ and $a_0$ for distribution function $f_{\rm
    Shu}(E,L)$ with  $v_{\rm circ}(R)=v_c\sqrt{1+ R_d/R}$. For
  description see Figure2.
\label{fig:f6}}
\end{figure}
We now repeat the same exercise for cases where the rotation curve 
is not flat. Specifically, we investigate the following two cases.
\begin{itemize}
\item Rising rotation curve: The circular velocity is assumed to be 
a power law with positive slope.
\be
v_{\rm circ}(R) &= & v_c (R/R_d)^{\beta} \ {\rm with}\  \beta=0.2 \\
\phi(R)  & = & v_c^2 (R/R_d)^{2\beta}/(2\beta) \ {\rm with}\  \beta=0.2 
\ee
\item Falling rotation curve: In this case we assume the 
potential to be a superposition of a point mass and a flat rotation
curve. The circular velocity decreases monotonically with radius 
but at large $R$ asymptotes to a constant value. 
\be
v_{\rm circ}(R) & = & v_c^2 (1 +\sqrt{R_d/R}) \\
\phi(R) & = & v_c^2 (\log(R/R_d) -R_d/R)
\ee
It should be noted 
that we do not study the
case of a rotation curve falling as a power law,
as the integral of the 
Shu distribution function over all space or more specifically 
the factor $g_K(a,R_g)$ becomes infinite.
\end{itemize}

The results for the above two rotation curves are shown in \fig{f3}, \fig{f4}, 
\fig{f5} and \fig{f6} and the fit parameters $c_1, c_2, c_3$ and $c_4$
are summarized in \tab{tb1}. 
It can be seen from \fig{f3} and \fig{f5} that the  $\Delta \Sigma$
curves follow an almost universal form. The functional
form differs only slightly from the case for a flat 
rotation curve. From \fig{f4} and \fig{f6}, it can be seen 
that the functional form given by \equ{rdelta1} and \equ{rdelta2}
provides a good fit for the dependence of $R_g^{\rm max}$ 
and $\Delta \Sigma^{\rm max}$ on $q$ and $a_0$ respectively. 
A slight residual dependence on $q$ of $\Delta \Sigma^{\rm max}$ 
is visible for falling rotation curve (lower right panel of \fig{f6}).
If we ignore the dependence of parameters $c1,c2,c3,c4$ on the 
rotation curve, i.e., $\beta$, the resulting surface density 
profiles will be slightly inaccurate.  
For $(a0,q)=(0.5,0.33)$, comparing 
$\Sigma_{\rm corr}(R_g)$ of $\beta=0$ with that of $\beta=0.2$,
we find 
that maximum deviation occurs at around $R \sim R_g^{\rm max}$ 
and is about 10\%.

\subsection{Case of $f_{\rm Dehnen,a}$ with a flat rotation curve}
\begin{figure}
 \centering \includegraphics[width=0.48\textwidth]{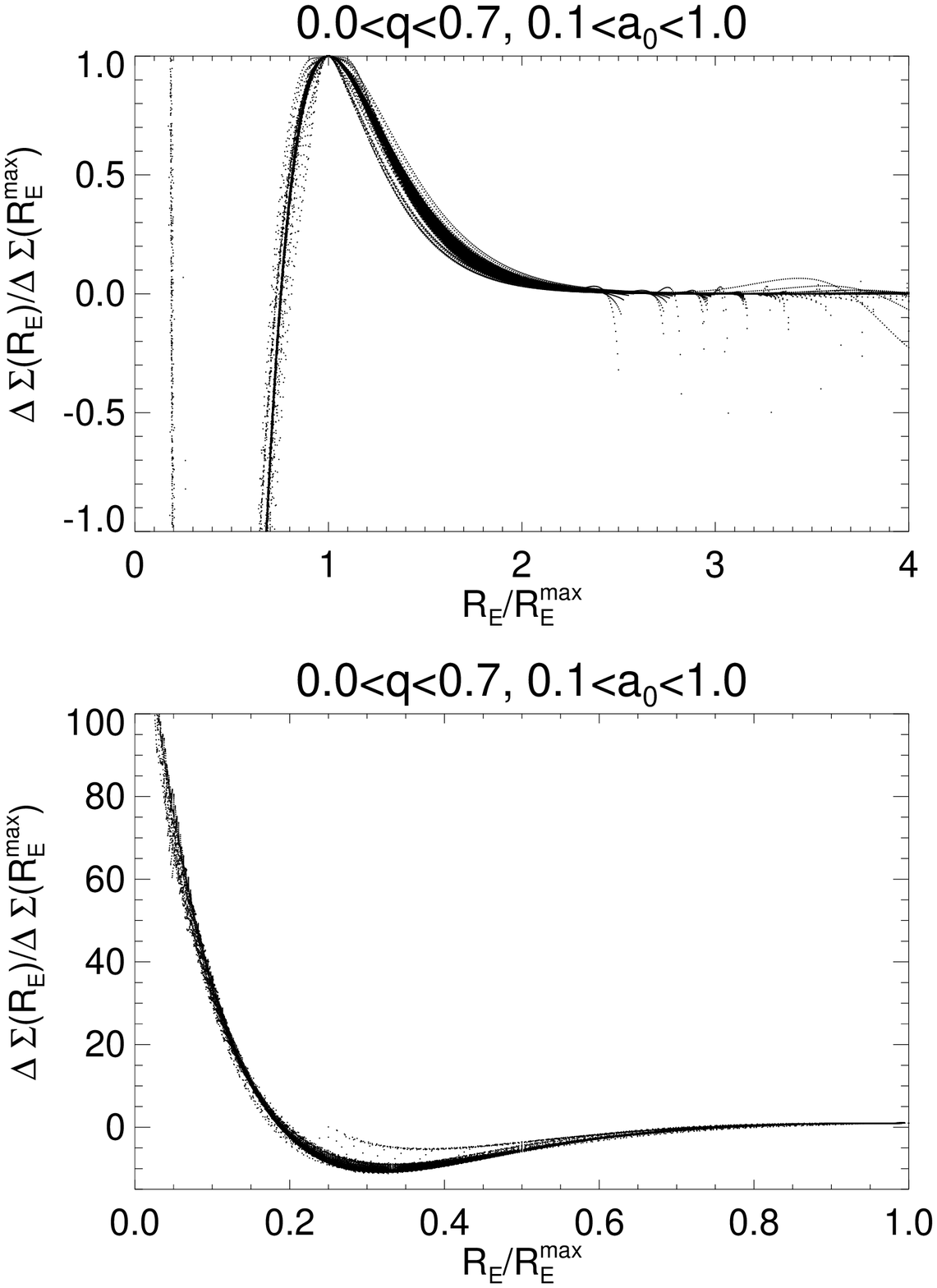}
\caption{ The correction factor $\Delta
  \Sigma$ for distribution function $f_{\rm Dehnen,a}(E,L)$ with
  $v_{\rm circ}(R)=v_c$. For  description see Figure1.
\label{fig:f7}}
\end{figure}

\begin{figure}
 \centering \includegraphics[width=0.48\textwidth]{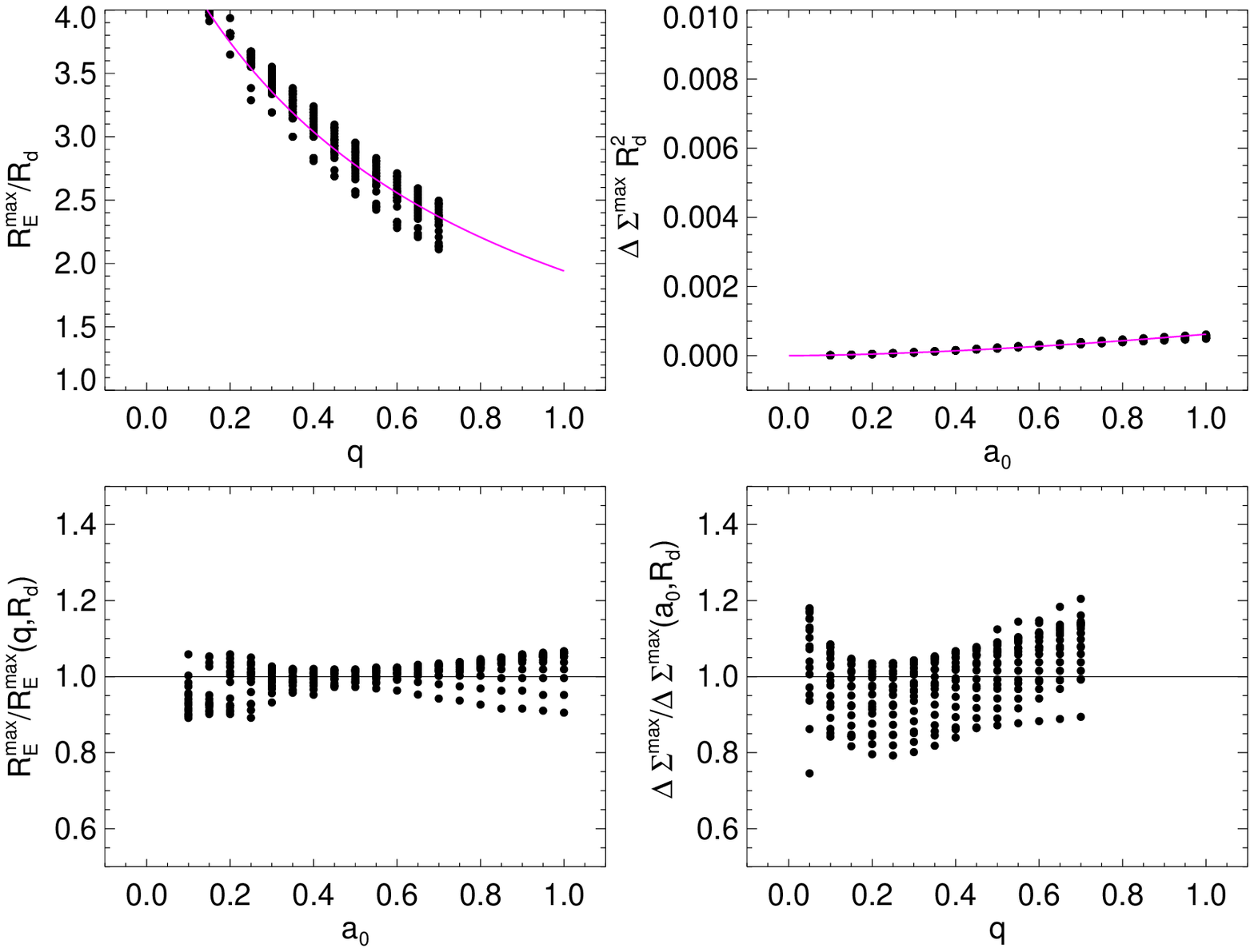}
\caption{ The dependence of $R_{g}^{\rm max}$ and $\Delta
  \Sigma^{\rm max}$ on $q$ and $a_0$ for distribution function $f_{\rm
    Shu}(E,L)$ with  $v_{\rm circ}(R)=v_c$. For
  description see Figure2.
\label{fig:f8}}
\end{figure}
Finally, we study the case of a distribution function different from 
Shu. We use the function $f_{\rm Dehnen,a}$ from
\citet{1999AJ....118.1201D}. For this, we only study the case of a flat rotation 
curve. It can be seen from \fig{f7} that the $\Delta \Sigma$ curves 
again follow an almost universal form. However, the form is different 
from that of $f_{\rm Shu}$. The main difference being that as $R_E$ 
approaches zero the curve moves upwards to large positive values.  
The functional form of $R_E^{\rm max}(q)$ 
and $\Delta \Sigma^{\rm max}(a_0))$ is the same as in $f_{\rm Shu}$ but 
with different values for constants, and the relationships are slightly less
accurate than for $f_{\rm Shu}$ (see \fig{f8}). Interestingly, 
$\Delta \Sigma^{\rm max}$ is much smaller than the case for Shu 
distribution function.

To summarize, we find that for a given distribution function and
rotation curve, the correction factor $\Sigma(R_c)$ 
required to reproduce exponential discs for different values 
of $a_0$ and $q$ follows a universal functional form to high accuracy.
The scale length $R_c^{\rm max}$ and  the amplitude of this 
function $\Delta \Sigma^{\rm max}$ has a simple dependence 
on $a_0$ and $q$ which can be parameterized in terms of four 
constant $c_1$, $c_2$, $c_3$ and $c_4$. 
The fact that an empirical formula based on the above methodology
can be determined for different rotation curves and even 
for a distribution function other than $f_{\rm Shu}$ is very 
promising. The reason that the methodology works for 
different cases is related somewhat to the following three facts 
\begin{itemize}
\item The functional form of the required target density is same 
for all cases.
\item We have  parameterized the velocity dispersion in a way 
such that its dependence on $R_c$ is same for all cases.
\item The integral equation governing the computation of $\Delta \Sigma$ 
is the same for all cases except for the Kernel function and its 
normalization function $g_K$.
\end{itemize}
Given an integral equation of the type in \equ{integralequ1}, the determination 
of $\Sigma(R)$ from $\Sigma(R_c)$ can be thought 
of as a convolution operation with a Kernel function $K(R,R_c)$. 
To first order, if the kernels are similar and of the same width, 
the effect of convolution should be insensitive to the 
shape of the kernel. 
For the cases studied here the kernel 
in general peaks at $R=R_c$ and then falls of sharply for both 
$R<R_c$ and $R>R_c$. The exact shape is governed primarily 
by the form of the distribution function. For a constant rotation 
curve, the width of the kernel is an exponentially decreasing function 
of $R_c$. The effect of a non-flat rotation 
curve is to simply modulate this functional dependence of width as a 
function of $R_c$.

\begin{figure}
 \centering \includegraphics[width=0.48\textwidth]{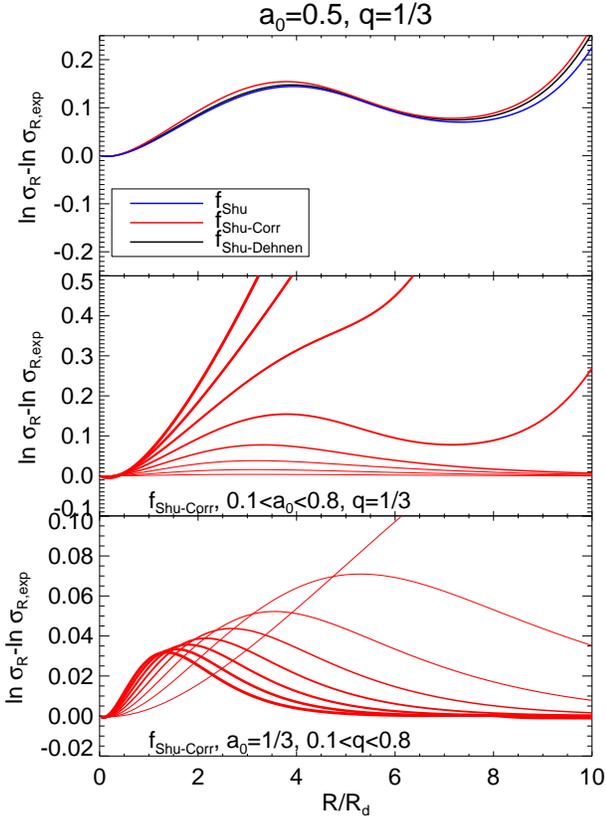}
\caption{ The logarithmic difference of the radial velocity dispersion
  profile $\sigma_R(R)$ with respect to the target velocity dispersion 
profile $\sigma_{R,{\rm exp}}(R)=v_c a_0\exp(-q R/R_d)$. 
In the top panel the three cases shown are for the Shu distribution function
with flat rotation curve, a) without the empirical formula ($f_{\rm Shu}$)
b) with the empirical formula ($f_{\rm Shu-Corr}$) c) with Dehnen's ansatz
($f_{\rm Shu-Dehnen}$). The bottom two panels are for $f_{\rm Shu-Corr}$ only.   
Here the different lines correspond to different values of $a_0$ and
$q$ and the thickness of the line is proportional to the value of $a_0$ 
and $q$. The values of $a_0$ and $q$ increase in steps of 0.1.
\label{fig:f9}}
\end{figure}

\begin{figure}
 \centering \includegraphics[width=0.48\textwidth]{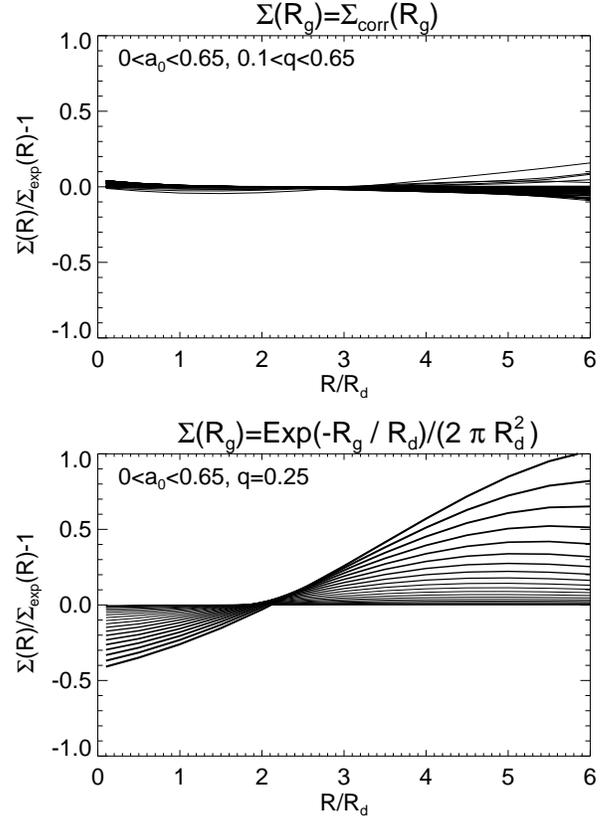}
\caption{ The fractional difference of the  surface density
  $\Sigma(R)$ from the target surface density for different 
  choices of $\Sigma(R_g)$.   The target surface density $\Sigma_{\rm exp}={\rm
    exp}(-R/R_d)/(2\pi R_d^2)$  is that of 
  an exponential disc. $\Sigma_{\rm corr}(R_g)$ 
  is the formula presented in this paper which approximately reproduces an
  exponential disc. Different lines correspond to
  different choices of $a_0$ and $q$. In the bottom panel $q$ is fixed 
  and the thickness of the lines is proportional to the value of $a_0$.
\label{fig:f10}}
\end{figure}

\begin{figure*}
 \centering \includegraphics[width=0.95\textwidth]{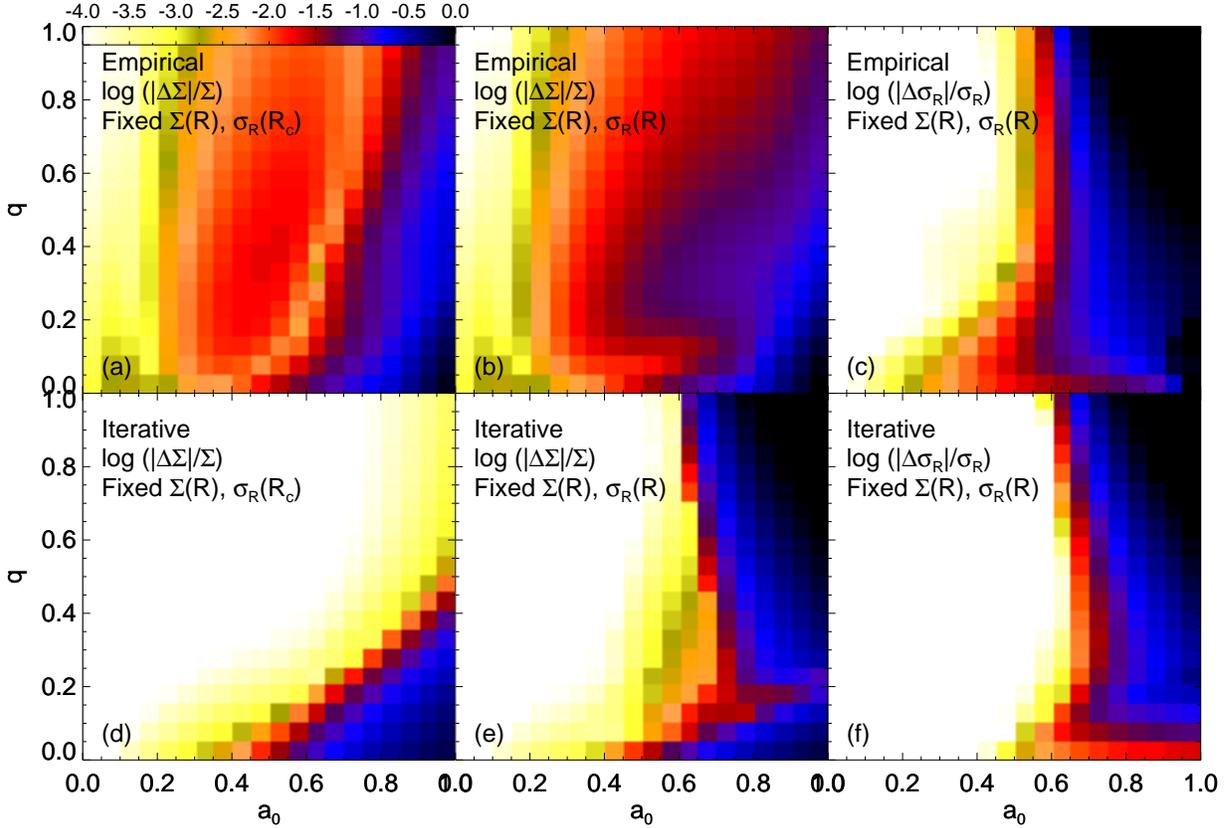}
\caption{ The mean absolute fractional difference of the final surface
  density $\Sigma(R)$ and velocity dispersion
  $\sigma_R(R)$ from the target surface density $\exp(-R/R_d)/(2 \pi
  R_d^2)$ and target velocity dispersion $v_c a_0\exp(-q R/R_d)$. 
  Shown is the case of $f_{\rm Shu}$ with flat rotation curve.
  The top panels are results from using the empirical formula proposed 
  in this paper. The bottom panels are results from using the
  iterative algorithm. The panels a and d 
  correspond to the case 
  where $\Sigma(R)$ and $\sigma_R(R_c)$ are specified to follow 
  exponential functional forms. The other panels are for the case 
  where $\Sigma(R)$ and $\sigma_R(R)$ are specified to follow 
  exponential functional forms.
  The mean difference was computed over equi-spaced bins in $0<R/R_d<5$,  
  and was not weighted for mass variation  else
  the contribution to the mean from large values of $R$ would be very small.
\label{fig:f11}}
\end{figure*}

\subsection{The case of velocity dispersion profiles that are
  exponential in radius}
\label{sec:velocity_disp}
Till now, we had assumed the $\sigma_R(R_c)$ profile to be fixed 
and exponential in $R_c$. However,   
this does not result in $\sigma_R(R)$ profiles that are strictly exponential 
in radius $R$. 
For $f_{\rm Shu}$ with flat rotation curve, the deviation 
of $\sigma_R(R)$ from that of 
an exponential form $\sigma_R(R)=v_c a_0 \exp(-q R/R_d)$ is shown in 
top panel of \fig{f9}. The deviation is very similar to 
the deviation noticed for surface densities, so the effect can 
be thought of as an increase in scale length of the exponential 
function governing the radial dependence of the dispersion profile.
The case of $f_{\rm Shu}$ both with and without the use 
of  empirical formula result in very similar  $\sigma_R(R)$ profiles. 
Additionally, a form of $f_{\rm Shu}$ with Dehnen's ansatz (see
\equ{dehnen_ansatz}) also gives a similar  $\sigma_R(R)$ profiles.   
This suggests 
that modifying the $F(R_c)$ part of the distribution function 
only has  a minor effect on the dispersion profiles. The primary 
quantity that determines the dispersion profile is the function 
$\sigma_R(R_c)$. It can be seen from \fig{f9} that the difference 
peaks at around $R/R_d=4$ then decreases and finally for large 
$R$ it starts to rise again. The bottom panel of \fig{f9} shows 
that the location of the 
peak primarily depends on $q$ (for small $q$, the peak is
at larger $R$) while the height depends on both $a_0$ and $q$. The larger 
the $a_0$ and smaller the $q$, the higher the peak. 
The rise at large $R$ is mainly 
dependent on $a_0$ and is stronger for large $a_0$. This 
is because the distribution of guiding center $R_g$ in a given 
annulus at large $R$ is bimodal. 
The main peak being from stars with $R_g \sim R $ but there is 
also a secondary peak from stars with very small $R_g$.
The secondary peak has a non-negligible contribution to the velocity 
dispersion as the corresponding  
$\sigma_R(R_g)$ is very large due to the exponential 
functional dependence. Unlike us, this rise at large $R$ is not visible 
in Figure-1a of \citet{1999AJ....118.1201D} 
and we believe that this is 
because in \citet{1999AJ....118.1201D}  
the convolution integral is not done over the full domain of $R_g$ 
but instead only around $R_g \sim R$. 

If one desires $\sigma_R(R)=v_{\rm circ}(R)a_0\exp(-q R/R_d)$
then one has to use the \citet{1999AJ....118.1201D} iterative solution 
to solve for both $\sigma_R(R_c)$ and  $\Sigma(R_c)$
simultaneously. We do this for the case of $f_{\rm Shu}$ 
with flat rotation curve. Doing so, we find that the 
$\Sigma_R(R_g)$ profile 
is almost unchanged from the case where $\sigma_R(R)$ was not
constrained. 
For $\sigma_R(R_g)$ we find that an empirical formula 
as given below can reproduce to a good accuracy 
the desired velocity dispersion profiles.
\be
\sigma_R^{\rm corr}(R_g)  & = & v_c a_0\exp(-q R_g/R_d) \times
\nonumber \\ 
& & \left(1-\frac{0.25 a_0^{2.04}}{q^{0.49}} f_{\rm  poly}(R_g
  q/R_d)\right), 
\label{equ:sigma_formula}
\ee
Here $f_{\rm poly}$ is an 11 degree polynomial having 
coefficients- $0.028476$, $-1.4518$, $12.492$, $-21.842$, $19.130$, $-10.175$
, $3.5214$, $-0.81052$, $0.12311$, $-0.011851$, $0.00065476$, $-1.5809
\times 10^{-5}$.

\section{Accuracy in reproducing the target density}
In the top panel of \fig{f10}, we plot the fractional difference of the 
final surface density $\Sigma(R)$, 
as compared to that of an exponential disc, 
using our empirical formula as given by \equ{density_formula} for the 
case of $f_{\rm Shu}$ with flat rotation curve. 
Results for a range of 
values of $a_0$ and $q$ are shown. For $R/R_d<5$, the difference 
is less than $10\%$.  The fit starts to 
deteriorate only for $R/R_d>5$.  
In general, as $a_0$ is increased, the fit progressively deteriorates.
For comparison, the bottom panel shows the fractional difference of 
$\Sigma(R)$ when a simple exponential form for $\Sigma(R_g)$ 
is adopted. The difference increases as $a_0$ is increased and is
negligible only for very small values of $a_0$. The improvement
offered by the empirical formula is clearly evident here.

The accuracy of the formula as a function of $a_0$ and $q$ can be 
better gauged in panels a and d of \fig{f11} . Here we plot the mean absolute fractional difference of surface density
$\Sigma(R)$ with respect to the target surface density $\Sigma_{\rm
  exp}(R)=\exp(-R/R_d)/(2\pi R_d^2)$ as a function of $a_0$ and
$q$. It can be seen that the proposed solution works quite well 
over most of the $(a_0,q)$ space except for very high values of
$a_0$. It should be noted that in the region $a_0>q+0.5$, 
lower right corner,  where the empirical formula fails  
the iterative solution also itself fails (see
panel d). So the actual 
inaccuracy of the formula is mainly confined to the region 
$(q>0.4,a_0>0.75)$.

We now study the accuracy of our analytic formula given by 
\equ{density_formula} and \equ{sigma_formula} for the case 
where both the target surface density and the velocity 
dispersion are specified to follow exponential forms.
The mean absolute fractional difference  of $\Sigma(R)$ and 
$\sigma_R(R)$ with respect to $\Sigma(R)=\exp(-R/R_d)/(2\pi R_d^2)$ and 
$\sigma_R=v_c a_0\exp(-q R/R_d)$ is shown in panels b and c of
\fig{f11}.  For comparison, the panels e and f show the results 
of the iterative algorithm.  
It can be seen that the formula works well in the range $0<a_0<0.5$,  
where the mean error is less than 1\%. However for $a_0>0.5$,  
the formula and the iterative solution both do not work well. 
It should be noted that the sharp transition 
at $a_0 \sim 0.5$ is due to the fact that we measure the 
difference over the range $0<R/R_d<5$. Extending the 
range would shift the transition value of $a_0$ to lower values 
and vice versa. In general, for a given $a_0$ the solution 
provided by the iterative method and the empirical formula are 
both accurate in reproducing target profiles for small values of $R$,  
they gradually become inaccurate at larger values of $R$. 
It can also be seen from the figure that for $a_0>0.5$,  
the empirical formula performs slightly better than the iterative
solution at reproducing 
the target surface densities. However, this is only at the expense of not 
matching the target velocity dispersion profiles as well.

We now discuss the possible causes for the failure of the iterative 
solution for large $a_0$. The iterative algorithm will only work 
correctly if the contribution to the integral for $\Sigma(R)$ 
or  $\sigma_R(R)$ (see Equation \ref{equ:integralequ1} and
\ref{equ:integralequ2}), is confined to a local region around $R_c \sim R$. 
In general, at large $R$ this condition is violated. 
The radius at which this violation 
occurs depends upon the choice of $a_0$, and is lower for larger 
values of $a_0$. 
This violation of the locality condition is stronger 
for $\sigma_R(R)$ profiles, as 
 $\sigma_R(R_c)$ increases exponentially with decrease in $R_c$. 
The rise in $\Delta \sigma_R/\sigma_R$ at large $R$ as discussed 
in \sec{velocity_disp} and \fig{f9} is also a 
manifestation of this effect.
This is the main reason that when the constraint of exponential 
radial dispersion profiles is applied the algorithm fails 
for $a_0>0.5$.


\begin{figure}
 \centering \includegraphics[width=0.48\textwidth]{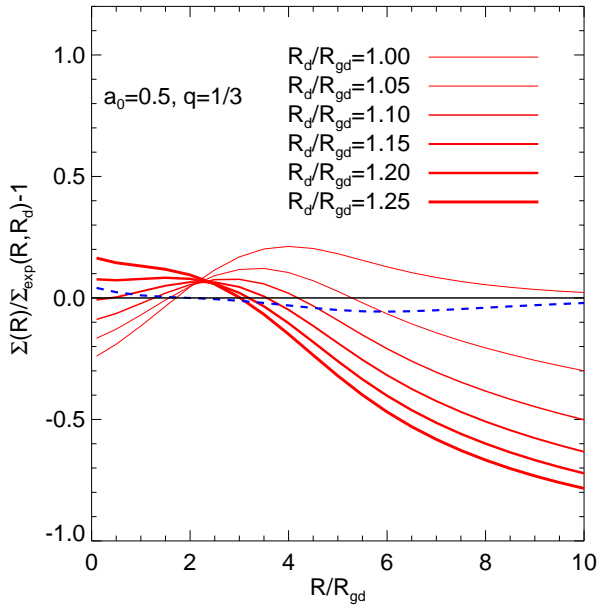}
\caption{ The fractional difference of the  surface density
  $\Sigma(R)$ from the target surface density, $\Sigma_{\rm exp}={\rm
    exp}(-R/R_d)/(2\pi R_d^2)$, for $\Sigma(R_g)={\rm
    exp}(-R_g/R_{gd})/(2\pi R_{gd}^2)$.  The thickness of the line 
  is proportional to the 
  value of $R_d$ used for $\Sigma_{\rm exp}$. 
  The dashed line is the result of using our empirical formula  
  with $R_{gd}=R_d$. 
\label{fig:f12}}
\end{figure}

\subsection{Comparison with other approximate solutions}
An alternative way to generate an exponential disc using 
a Shu type distribution function was proposed in 
\citet{2010MNRAS.401.2318B}. They note that the effect 
of warming up the distribution function is to expand the disc, 
and this can be thought of as an increase in scale length of the
disc. So, one way to take this into account is to start with a slightly
smaller scale length when specifying $\Sigma(R_g)$. There are two
problems with this approach. First, the
factor by which the scale length will have to be reduced will depend 
upon the values of $a_0$ and $q$. So the solution is no simpler 
than what we propose. Secondly, we find that 
even if one has the correct factor, the solution is less than 
optimal. To show this, we plot in \fig{f12} the fractional difference 
of the surface density $\Sigma(R)$ obtained using
$\Sigma(R_g)={\rm exp}(-R_g/R_{gd})/(2\pi (R_{gd})^2)$ 
from an exponential surface density with different values of $R_d$.
We set $a_0=0.5$ and $q=0.33$ which is typical of old thin disc stars.
The case considered is that of $f_{\rm Shu}$ with flat rotation curve.
It can be seen that increasing $R_d$ improves the solution slightly 
for $R/R_{gd}<5$, but at the expense of deteriorating it for large $R$.  
Overall, the quality of the fit is quite poor.
For comparison, the dashed line is the result 
using our empirical formula  with $R_{gd}=R_{d}$,  
which clearly performs better.

\begin{figure}
 \centering \includegraphics[width=0.48\textwidth]{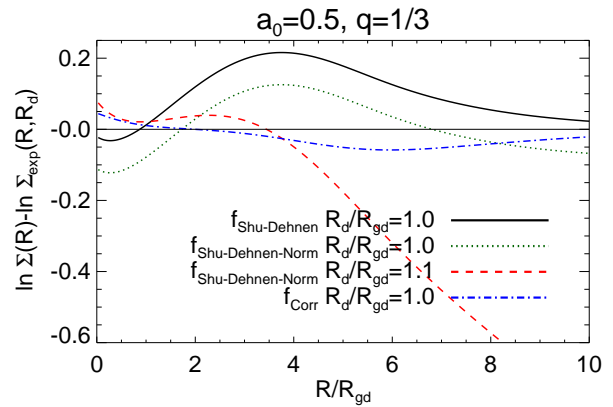}
\caption{ The logarithmic difference of the  surface density
  $\Sigma(R)$ from the target surface density, $\Sigma_{\rm exp}={\rm
    exp}(-R/R_d)/(2\pi R_d^2)$, for $\Sigma(R_g)={\rm
    exp}(-R_g/R_{gd})/(2\pi R_{gd}^2)$. Shown are the results for 
the Dehnen's ansatz with different scale length. The results 
of our empirical formula are also shown alongside. 
\label{fig:f13}}
\end{figure}

One might argue that the ansatz of \citet{1999AJ....118.1201D}
\be
f_{\rm
  Shu-Dehnen}(E,L)=\frac{\gamma(R_g)\Sigma(R_g)}{2\pi\sigma_R^2(R_g)}\exp\left(\frac{E_c(R_g)-E}{\sigma_R^2(R_g)}\right)
\label{equ:dehnen_ansatz}
\ee
actually produces discs 
that are more close to an exponential form, 
and so applying the \citet{2010MNRAS.401.2318B} scale length reduction to it 
might produce even better agreement. We now investigate this issue.
Firstly, the above distribution function is not normalized.
If $\sigma_R(R_g)$ is parameterized in terms $a_0$ and $q$ as before 
then it has a normalization that depends upon the choice of 
$a_0$ and $q$, so the full distribution function is not purely 
analytical anymore. 
On top of that as said earlier, the
factor by which the scale length will have to be reduced will also 
depend upon the values of $a_0$ and $q$. We assume it to be 
10\% for the time being.
In \fig{f13}, we show the fractional difference of 
surface density from the target density for the case of $f_{\rm
  Shu-Dehnen}$ with flat rotation curve. 
Results with normalization (denoted by $f_{\rm Shu-Dehnen-Norm}$) 
and  with \citet{2010MNRAS.401.2318B} scale length reduction are also shown 
alongside. 
It can be seen that after normalizing  
the Dehnen's ansatz does produce discs that match  the target density better, 
i.e., the range of deviation is smaller 
than the thin red line corresponding to $R_g/R_d=1$ in \fig{f12}. 
After applying the \citet{2010MNRAS.401.2318B} scale length 
reduction  the agreement gets even better, but only in regions with 
$R/R_d<4$. For large $R$, there is still a wide discrepancy. 
In comparison, it can be seen that the formula proposed in this paper 
(dashed blue line) is still superior while being simpler and 
fully analytic.

\section{Discussion}
In this paper, we have presented an empirical formula for
Shu-type distribution functions  
which can reproduce with good accuracy a disc with an
exponential surface density profile. 
This should be useful in constructing equilibrium N-body models of
disc galaxies
\citep{1995MNRAS.277.1341K,2007MNRAS.378..541M,2005ApJ...631..838W}
where such functions are employed. 
It can also be employed in synthetic Milky Way
modelling codes like TRILEGAL \citep{2005A&A...436..895G}, 
Galaxia \citep{2011ApJ...730....3S} and  
Besancon \citep{2003A&A...409..523R}.
Finally, the most important use of our new formalism is for MCMC fitting 
of theoretical models to large data sets, e.g., 
the Geneva-Copenhagen \citep{2004A&A...418..989N} , 
RAVE \citep{2006AJ....132.1645S} and SEGUE stellar surveys
\citep{2009AJ....137.4377Y}. 
Here, separable analytic approximations of this kind are required
to make the problem tractable (Sharma et al 2013, in preparation).

One of our main finding is that the part of the distribution function 
that determines the target surface density can be written 
as the target density plus a correction term which approximately 
follows a unique functional form. So, the problem reduces to determining the 
scale and the amplitude of this correction function and 
their dependence on  $a_0$ 
(ratio of central velocity dispersion to circular velocity) and $q$ 
(a parameter controlling the gradient of the dispersion as a 
function of radius). 
The problem is further simplified as we find that 
the scale is primarily a function of only $q$ and the amplitude is 
a primarily a function of only $a_0$. These 
findings are valid for not only flat rotation curves but 
also for cases with rising and falling rotation curves. Additionally,  
it was also found to be valid for one of the \citet{1999AJ....118.1201D} 
distribution functions. 
An extension of the formula to reproduce 
velocity dispersion profiles that are exponential in radius was 
also found to work well.
This suggests  that the methodology presented
can, in principle, be also applied to other disc distribution functions 
proposed by \citet{1999AJ....118.1201D} and 
\citet{2010MNRAS.401.2318B,2012MNRAS.426.1328B}
which are conceptually similar.

\section*{Acknowledgments} 
We are thankful to the anonymous referee who among other things 
motivated us to generalize our results. We are also thankful 
to James Binney for his comments that led to
a substantial improvement in the manuscript.
JBH is funded through a Federation Fellowship from the Australian 
Research Council (ARC). SS is funded through ARC DP grant 120104562
which supports the HERMES project.

\appendix
\section{Integral equation for one of the Dehnen distribution functions}
We consider here the distribution function 
\be
f_{\rm Dehnen,a} & = &\frac{F(R_E)}{\sigma^2_R(R_E)}\exp\left(-\frac{E-E_c(L)}{\sigma^2_R(R_E)}\right)
\ee
from \citet{1999AJ....118.1201D}. For $f_{\rm Dehnen,a}$ one can write the probability distribution in $(R,R_E,R_g)$
space as 
\be
P(R,R_E,R_g)dR dR_E dR_g &= &2\pi f(E,L)dL dR 2d |v_R| \\
P(R,R_E,R_g) &= &2\pi f(E,L) \frac{dL}{dR_g} \frac{dE}{dR_E}
\frac{2}{\sqrt{2(E-\Phi(R))-L^2/R^2}}
\ee
To proceed further we assume a flat rotation curve.
Using $E=\Phi_{\rm eff}(R_E,R_E)$, $E_c(R_g)=\Phi_{\rm
  eff}(R_g,R_g)$, $dL/dR_g=\frac{2 v_{\rm circ}(R_g)}{\gamma^2(R_g)}$, $dE/dR_E=\frac{2
  v_{\rm circ}^2(R_E)}{\gamma^2(R_E)R_E}$, $\gamma^2(R_g)=\gamma^2(R_E)=2$ we get
\be
P(R,R_E,R_g) &= & \frac{4 \pi F(R_E)}{a^2(R_E)} \frac{R_E^{-(1+\frac{1}{a^2})}
  R_g^{\frac{1}{a^2}}}{\sqrt{1-2\ln(R/R_E)-R_g^2/R^2}} 
\ee
Substituting $\frac{R_g}{R\sqrt{1-2\ln(R/R_E)}}=x$ and 
\be
K(R,R_E)=(R/R_E)^{(1+\frac{1}{a^2})}(1-2\ln(R/R_E))^{\frac{1}{2a^2}}
\ee
this simplifies to
\be
P(R,R_E,x) &= & \frac{4 \pi F(R_E)}{a^2(R_E)} K(R,R_E) \frac{x^{\frac{1}{a^2}}}{\sqrt{1-x^2}} 
\ee
Integrating over $x$ from 0 to 1 we get
\be
P(R,R_E) &= & \frac{4 \pi F(R_E)}{a^2(R_E)} K(R,R_E) h(a)   
\ee
where
\be
h(a)=\int_{0}^{1}\frac{x^{1/a^2}}{\sqrt{1-x^2}}=\frac{\sqrt{\pi}}{2}\frac{\Gamma(\frac{1}{2a^2}+\frac{1}{2})}{\Gamma(\frac{1}{2a^2}+1)}=\frac{\sqrt{\pi}}{2}\frac{\Gamma(c+\frac{1}{2})}{\Gamma(c+1)} \ {\rm with}\ c=\frac{1}{2a^2}
\ee
Using $\int P(R,R_E) dR= 2\pi R_E \Sigma(R_E)$ one can write
$F(R_E)$ in terms of $\Sigma(R_E)$ to get
\be
F(R_E) &= & \frac{ \Sigma(R_E) a^2(R_E)}{2 g_K(a) h(a)}
\ee
and then 
\be
P(R,R_E) &= & \frac{2 \pi \Sigma(R_E)}{g_K(a)} K(R,R_E)
\ee
Here
\ben
g_K(a) & = & \frac{1}{R_E} \int K(R,R_E) dR = \left(\frac{e}{1+c}\right)^{1+c} \frac{\Gamma(1+c)}{2} \ {\rm with}\ c=\frac{1}{2a^2}\\
& \approx & \sqrt{\frac{\pi}{2(c+1)}} \ {\rm for\ large\ c}. 
\een

\bibliographystyle{apj} 

\end{document}